\documentclass[aps,twocolumn]{revtex4}
\usepackage{graphics}
\usepackage{dcolumn}
\usepackage{bm}
\usepackage{amsmath}
\usepackage{newlfont}
\usepackage{amssymb}
\usepackage{amsfonts}
\usepackage{amsmath}
\usepackage{graphicx}
\usepackage{bm}

\def \be{\begin{equation}}
\def \ee{ \end{equation} }

\begin{document}

\title{Locally Accessible Information of Multisite Quantum Ensembles Violates Monogamy}




\author{Aditi Sen(De) and Ujjwal Sen}

\affiliation{Harish-Chandra Research Institute, Chhatnag Road, Jhunsi, Allahabad 211 019, India}


\begin{abstract}
Locally accessible information is a useful information-theoretic physical quantity of an ensemble of multiparty quantum states. 
We find it has properties akin to quantum  as well as classical correlations of single multiparty quantum states. 
It satisfies monotonicity under local quantum operations and classical communication. However we show that it does not follow monogamy, 
an important property usually satisfied by quantum correlations,
and actually violates any such relation to the maximal extent.  
Violation is obtained even for locally indistinguishable, but globally orthogonal, multisite ensembles. 
The results assert that while single multiparty quantum states
are monogamous with respect to their shared quantum correlations, ensembles of multiparty quantum states may not be so. 
The results have potential implications for quantum communication systems.
\end{abstract}

\maketitle

The science of quantum information \cite{NC} had its origins in thermodynamics \cite{thermodynamics} and foundations of quantum mechanics \cite{Bell}, 
and it has since been successfully applied 
to computation \cite{compu} and communication sciences \cite{comm}, and has also found interesting links to many-body physics \cite{many-body}.  
The field typically deals with a system of many different subsystems,  and 
quantum correlations \cite{HHHHRMP}, in its multifarious versions, between the different subsystems form the backbone of these applications. 
The excitement in quantum information is even more because numerous laboratories around the globe can realize entanglement in 
a variety of different physical systems.
While these different measures of quantum correlation have quite diverse motivations for their introduction and have otherwise dissimilar regions of utility, 
they do share certain intuitively satisfying criteria \cite{HHH-limits}.

An important property that is expected to be satisfied by quantum correlation measures is 
monotonicity (more precisely, non-increasing), 
in some form,  under local quantum 
operations and classical communication (LOCC). 

Another prominent criterion that is usually expected from a measure of quantum correlation is 
the so-called ``monogamy''. Such a property is expected to be 
active in any measure of quantum correlation for any quantum state of a multiparty system, where each of the parties possess one of the subsystems. 
Monogamy requires that for any quantum state 
of a multiparty system, if two parties (i.e. the subsystems in possession of the respective parties)
are highly correlated according to a certain measure of quantum correlation, then these parties would not have a substantial amount of that quantum correlation 
measure with any other third party (subsystem). See  \cite{monogamy, monogamy-also}.

None of these dual fundamental properties -- monogamy and monotonicity -- are expected to be satisfied by a classical correlation measure of a multiparty quantum state. 

The concept of accessible information of a quantum ensemble is one that predates the usually accepted beginnings of the field of quantum information, and is 
defined as the maximal amount of classical information that one can obtain from a quantum ensemble by using arbitrary quantum measurements. 
The ``Holevo bound'' for accessible information 
obtained about 40 years earlier, provides us with an important piece of information: bit per qubit, i.e. the amount of classical information that can be 
incorporated in a two-dimensional quantum system (qubit) is one bit (binary digit) \cite{Holevo}. 

A related concept is that of locally accessible information and is the maximal amount of classical information that one can obtain from a 
multiparty quantum ensemble, when only \emph{local} quantum measurement strategies (LOCC-based measurement strategies) 
are allowed -- quantum measurements at all the subsystems and 
classical communication between them. 
The first result in this direction was very surprising: It was shown that there exists   sets of orthogonal product states that are not locally distinguishable --
``quantum nonlocality without entanglement''
\cite{nlwe,UPB1, UPB-rest}. On the other hand, it turned out that any two orthogonal pure states of arbitrary dimensions and number of parties are locally distinguishable
irrespective of its quantum correlation content \cite{two-state}. Furthermore, an example was given of an ensemble of bipartite quantum states
which is locally distinguishable, but on \emph{reducing} its average quantum correlation content, the ensemble becomes locally indistinguishable -- ``more 
nonlocality with less entanglement'' \cite{morenon}. See also \cite{Peres-Wootters-and} in this regard.

Notwithstanding these counterintuitive properties of locally accessible information, the latter does have a certain amount of direct 
proportionality with quantum correlation, as was found in 
\cite{directly-proportional}.
The Holevo bound states that using a quantum ensemble of a system of \(n\) qubits, it is not possible to send more than \(n\) bits of classical information. 
Ref. \cite{directly-proportional} demonstrates a local version of this result: Using a quantum ensemble of a bipartite system of 
\(n\) qubits, it is not possible to send more than \(n-\overline{E}\) bits of classical information. Here \(\overline{E}\) denotes the average, over the ensemble,
of an entanglement measure that satisfies certain basic postulates. 

For an ensemble of multipartite quantum states, its locally accessible information is an important and useful 
information-theoretic physical quantity
of the ensemble.
In this paper, we find that locally accessible information possesses properties that are similar to those of quantum 
as well as of classical correlations of single multisite quantum states. 
In particular, it certainly satisfies the monotonicity postulate: locally accessible
information of an ensemble of quantum states is nonincreasing under LOCC on the ensemble. 
However, we show here that the same physical quantity is polygamous, even for multiparty quantum ensembles that are LOCC indistinguishable.
We also find locally distinguishable ensembles that violate any monogamy relation to the 
maximal extent.

For an ensemble of multiparty quantum states, shared between the \(N+1\) parties, \(A\), \(B_1\), \(B_2\), ... \(B_N\),
the accessible information \(I_{acc}^{LOCC, A:B_i}\) of the ensemble, when we have access to only the parties \(A\) and \(B_i\) (\(i=\) 
any one of \(1,2, \ldots, N\)), and 
when only LOCC is allowed between \(A\) and \(B_i\), 
monogamy would dictate that \cite{monogamy}
\begin{equation}
 \sum_{i=1}^{N} I_{acc}^{LOCC, A:B_i} \leq I_{acc}^{LOCC, A:B_1 B_2 \ldots B_N}. 
\end{equation}
We show that there exist locally indistinguishable (but globally orthogonal) ensembles that violate this relation.

Moreover, note that for an arbitrary ensemble shared between the \(N+1\) parties,
we certainly have \(I_{acc}^{LOCC, A:B_i} \leq \log_2 \Gamma\), where \(\Gamma\) denotes the 
cardinality of the ensemble \cite{logbase2}. 
Monogamy of locally accessible information would provide a bound on \(\sum_{i=1}^{N} I_{acc}^{LOCC, A:B_i}\) that is strictly lower than 
\(N\log_2\Gamma \).
Quite contrarily, we show that there exists ensembles of multiparty quantum states, for which the sum 
\(\sum_{i=1}^{N} I_{acc}^{LOCC, A:B_i}\) violates monogamy to the maximal extent -- it attains the  value \(N\log_2\Gamma\).

The results have potential applications in quantum communication networks and in dealing with the information dynamics in quantum computation devices, especially 
in distributed quantum computing.
Moreover, since we demonstrate the results for arbitrary \(N\), the violation of monogamy obtained, persists for macroscopic systems. This, 
in particular, has potential implications for the way in which the quantum-to-classical transition is currently being viewed.
Despite its importance in the field of secret quantum communication, 
quantum correlations of an ensemble of multiparty quantum states is not as yet a well-developed field, and the results obtained has potential of producing an 
axiomatic formalism for such a quantity (see \cite{amader-q} in this regard), just like the one existing for quantum correlations of a single quantum state.

Let us begin by briefly presenting a formal definition of accessible information and locally accessible information. 
Suppose that an observer, Alice, obtains the classical message \(i\), and it is known 
that 
the message appears with probability \(p_i\). 
She wants to send it to another observer, Bob. 
To this end, Alice encodes the information \(i\) in a quantum state \(\rho_i\), and sends 
the quantum state to Bob. 
Bob receives the ensemble \(\{p_i, \rho_i\}\), and wants to obtain as much information as possible about \(i\). 
He performs a quantum measurement, that produces the result \(m\), with probability \(q_m\). Let the corresponding post-measurement ensemble be 
\(\{p_{i|m}, \rho_{i|m}\}\). The classical information that is gathered about the index \(i\) by the quantum measurement, 
can be quantified by the mutual information between the 
message index \(i\) and the measurement outcome \(m\) \cite{Chennai}:
\begin{equation}
I(i:m)= H(\{p_i\}) - \sum_m q_m H(\{p_{i|m}\}).
\end{equation}
Here \(H(\{r_\alpha\}) = -\sum_\alpha r_\alpha\log_2r_\alpha\) is the Shannon entropy of the probability distribution \(\{r_\alpha\}\).
Note that the mutual information can be seen as the difference between the initial disorder and the (average) final disorder. 
Bob will be interested to obtain the maximal information, which is the maximum of \(I(i:m)\) over all measurement strategies. This 
quantity is called the accessible information,
\begin{equation}
I_{acc} = \max I(i:m),
\end{equation} 
where the maximization is over all measurement strategies. 
The Holevo bound provides a universal upper bound \cite{Holevo} (see also \cite{Holevo-etc, directly-proportional}) on this quantity: 
\begin{equation}
I_{acc}\left(\{p_i, \rho_i\}\right) \leq \chi\left(\{p_i, \rho_i\}\right) \equiv S\left(\overline{\rho}\right) - \sum_i p_i S\left(\rho_i\right).
\end{equation}
Here \(\overline{\rho} = \sum_ip_i\rho_i\) is the average ensemble state, and 
\begin{equation}
\label{snajh-ekhon-Sushanta-dada-r sathhey-katha-bolchhe}
S(\varsigma)= - \mbox{tr}(\varsigma \log_2 \varsigma)
\end{equation} 
is the von Neumann entropy of the quantum state \(\varsigma\). 
The bound is universal in the sense that it is valid for arbitrary ensembles.
A weaker version of this result states that the accessible information for an ensemble of \(n\) qubit states is bounded above by \(n\) bits.

There also exists a universal lower bound on accessible information, and is given by \cite{JRW, JRW-also}
\begin{equation}
 I_{acc}\left(\{p_i, \rho_i\}\right) \geq Q(\overline{\rho}) - \sum_i p_i Q(\rho_i),
\end{equation}
where the ``subentropy'' \(Q\) is given by 
\begin{equation}
Q(\varsigma) =  - \sum_{k} \prod_{l\neq k} \frac{\lambda_k}{\lambda_k - \lambda_l} \lambda_k
\log_{2} \lambda_k,
\end{equation}
with \(\lambda_k\)'s being the eigenvalues of the state \(\varsigma\), and where one must consider the limit as the eigenvalues become equal, in the 
degenerate case.

To arrive at the concept of locally accessible information, let us again suppose that Alice has a message \(i\),
and that again it is known that the message happens with probability \(p_i\). But now, Alice encodes the message \(i\) in a \emph{bipartite} quantum state 
\(\varrho_i\). She sends one part of the bipartite state to an observer called Bob\(_1\) (\({\cal B}_1\)), and the other part to an observer called Bob\(_2\) 
(\({\cal B}_2\)). 
The Bobs therefore receive the ensemble \(\{p_i, \varrho_i^{{\cal B}_1{\cal B}_2}\}\), and their task is to gather as much information as possible about the 
index \(i\), by using only LOCC.
The maximal mutual information in this case is the locally accessible information,
\begin{equation}
I_{acc}^{LOCC} = \max I(i:m),
\end{equation} 
where the maximization is now over all LOCC-based measurement strategies. 
A 
universal upper bound on the locally accessible information is given by 
\cite{directly-proportional}
\begin{eqnarray}
I_{acc}^{LOCC}\left(\{p_i, \varrho_i^{{\cal B}_1{\cal B}_2}\}\right) \leq \chi^{LOCC}\left(\{p_i, \varrho_i^{{\cal B}_1{\cal B}_2}\}\right) 
                                                           \phantom{S\left(\overline{\varrho}^{{\cal B}_1}\right)} \nonumber \\
\phantom{S\left(\overline{\varrho}^{{\cal B}_1}\right)} \equiv S\left(\overline{\varrho}^{{\cal B}_1}\right) + S\left(\overline{\varrho}^{{\cal B}_2}\right) - 
\max_{k=1,2}\sum_i p_i S\left(\varrho_i^{{\cal B}_k}\right).
\end{eqnarray}
 Here \(\overline{\varrho}^{{\cal B}_1} = \mbox{tr}_{{\cal B}_2} \sum_i p_i \varrho_i^{{\cal B}_1{\cal B}_2}\), and similarly for \(\overline{\varrho}^{{\cal B}_2}\).
Also, \(\varrho_i^{{\cal B}_1} = \mbox{tr}_{{\cal B}_2} \varrho_i^{{\cal B}_1{\cal B}_2}\), and 
similarly for \(\varrho_i^{{\cal B}_2}\). 
Again, a weaker version of this result is available, which states that the locally accessible information of a bipartite ensemble of \(n\) qubits is bounded above by 
\(n - \overline{E}\), where \(\overline{E} = \sum_ip_iE\left( \varrho_i^{{\cal B}_1{\cal B}_2} \right)\) is the average entanglement \(E\) of the ensemble states. 
Here \(E\) is any measure of bipartite entanglement that satisfies \(E(\zeta^{{\cal B}_1{\cal B}_2}) \leq \max_{k=1,2} S\left(\zeta^{{\cal B}_k}\right)\) for 
all bipartite quantum states \(\zeta^{{\cal B}_1{\cal B}_2}\), where 
\(\zeta^{{\cal B}_1} = \mbox{tr}_{{\cal B}_2} \zeta^{{\cal B}_1{\cal B}_2}\), and similarly for \(\zeta^{{\cal B}_2}\).

Similarly, as in the case of accessible information with global operations, there also exist a universal lower bound 
on locally accessible information, and is given by \cite{amader-lower, Winter-lower}
\begin{eqnarray}
I_{acc}^{LOCC}\left(\{p_i, \varrho_i^{{\cal B}_1{\cal B}_2}\}\right) \geq \Lambda^{LOCC}\left(\{p_i, \varrho_i^{{\cal B}_1{\cal B}_2}\}\right) 
                                                           \phantom{S\left(\overline{\varrho}^{{\cal B}_1}\right)} \nonumber \\
\phantom{S\left(\overline{\varrho}^{{\cal B}_1}\right)} \equiv Q_L\left(\overline{\varrho}^{{\cal B}_1{\cal B}_2}\right)  
- \sum_i p_i Q_L\left(\varrho_i^{{\cal B}_1{\cal B}_2}\right).
\end{eqnarray}
Here \(\overline{\varrho}^{{\cal B}_1{\cal B}_2} = \sum_ip_i \varrho_i^{{\cal B}_1{\cal B}_2} \), and the ``local subentropy'' \(Q_L\)
is given by 
\begin{equation} 
Q_L(\zeta) = -   d_{{\cal B}_1} d_{{\cal B}_2} \int d\alpha d \beta \langle \alpha| \langle\beta| \zeta 
|\alpha \rangle  |\beta \rangle 
 \log_2   \langle \alpha|  \langle \beta| \sigma
|\alpha \rangle  |\beta \rangle,
\end{equation}
for a bipartite state \(\zeta\) of dimensions \(d_{{\cal B}_1} \otimes d_{{\cal B}_2}\).

With these concepts in hand, we will now probe the status of monogamy for locally accessible information.

\textbf{Case I.} We begin by considering the following set of three-qubit Greenberger-Horne-Zeilinger (GHZ) states \cite{GHZ}:
\begin{eqnarray}
|\psi_0^+\rangle_{AB_1B_2} &=& \frac{1}{\sqrt{2}}(|000\rangle + |111\rangle), \nonumber \\
|\psi_0^-\rangle_{AB_1B_2} &=& \frac{1}{\sqrt{2}}(|000\rangle - |111\rangle).
\end{eqnarray}
Therefore, \(A\) and \(B_1\) possess the quantum state 
\begin{equation}
\frac{1}{2}(|00\rangle\langle00| + |11\rangle\langle11|),
\end{equation}
irrespective of whether the three parties, \(A\), \(B_1\), and \(B_2\) share the state \(|\psi_0^+\rangle\) or \(|\psi_0^-\rangle\). 
Consequently, 
\begin{equation}
 I_{acc}^{LOCC,A:B_1} =0.
\end{equation}
Similarly, \( I_{acc}^{LOCC,A:B_2}\) is also vanishing, so that 
\begin{equation}
 I_{acc}^{LOCC,A:B_1} + I_{acc}^{LOCC,A:B_2} =0
\end{equation}
in this case. 
And, \(I_{acc}^{LOCC,A:B_1B_2} =1\) here, so that monogamy is satisfied in this case.
The two GHZ states considered are distinguishable by LOCC between \emph{all the three parties}. However,  the 
complete orthogonal basis of GHZ states spanning the three-qubit Hilbert space, which is locally indistinguishable, also satisfies (actually saturates) 
the monogamy relation.

\textbf{Case II.}
In complete contrast to the previous case, in this case study, we provide three concrete ensembles, each of which contains a 
different amount of average entanglement, such that each of them will violate any monogamy relation to the maximal extent. Each 
of the examples are three-qubit ensembles of two elements each. Let the three observers be again called 
\(A\), \(B_1\), and \(B_2\).
The maximal value that \(I_{acc}^{LOCC,A:B_1}\) can attain for a two element ensemble is unity. So is the case for \(I_{acc}^{LOCC,A:B_2}\). The first ensemble 
\(({\cal E}_1)\)
consists of the GHZ states 
\begin{eqnarray}
|\psi_0^+\rangle_{AB_1B_2} &=& \frac{1}{\sqrt{2}}(|000\rangle + |111\rangle), \nonumber \\
|\psi_3^+\rangle_{AB_1B_2} &=& \frac{1}{\sqrt{2}}(|011\rangle - |100\rangle).
\end{eqnarray}
Forgetting about \(B_2\), the ensemble consists of the states
\begin{eqnarray}
 \rho_{0+}^{AB_1} = \frac{1}{2}(|00\rangle \langle 00| + |11\rangle\langle11|), \nonumber \\
 \rho_{1+}^{AB_1} = \frac{1}{2}(|01\rangle \langle 01| + |10\rangle\langle10|).
\end{eqnarray}
This ensemble can be exactly distinguished by LOCC between \(A\) and \(B_1\), by  measurement in the computational basis at \(A\) and \(B_1\) and 
communication of the results (say, by a phone call), so that 
\begin{equation}
 I_{acc}^{LOCC,A:B_1}({\cal E}_1) = 1.
\end{equation}
The ensemble \({\cal E}_1\) is invariant under a swap operation between \(B_1\) and \(B_2\), so that we also have 
\begin{equation}
 I_{acc}^{LOCC,A:B_2}({\cal E}_1) = 1.
\end{equation}
Consequently, we have 
\begin{equation}
 I_{acc}^{LOCC,A:B_1}({\cal E}_1) + I_{acc}^{LOCC,A:B_2}({\cal E}_1) = 2,
\end{equation}
and 2 is the maximal value that the sum on the left-hand-side can attain (for an arbitrary two-element
ensemble), because the individual algebraic maxima are unity. 
This therefore is a violation of any monogamy relation that one can envisage for locally accessible information. 
The states in the ensemble \({\cal E_1}\) are genuinely multiparty entangled \cite{genuine-multi}. However, the violation of monogamy is not related 
to this fact, as borne out by the next two ensembles. Let us therefore consider the ensemble \({\cal E}_2\) that consists of the states
\begin{eqnarray}
|\psi^+\rangle_{AB_1B_2} &=& |0\rangle\frac{1}{\sqrt{2}}(|00\rangle + |11\rangle), \nonumber \\
|\psi^-\rangle_{AB_1B_2} &=& |1\rangle\frac{1}{\sqrt{2}}(|00\rangle - |11\rangle).
\end{eqnarray}
Again we have 
\begin{equation}
 I_{acc}^{LOCC,A:B_1}({\cal E}_2) + I_{acc}^{LOCC,A:B_2}({\cal E}_2) = 2,
\end{equation}
and again 2 is the maximal value that the sum on the left-hand-side can attain, because the individual algebraic maxima are unity. 
The states of this ensemble are still entangled, although not genuinely multisite entangled. The third and last ensemble (\({\cal E}_3\))
consists of the states 
\(|000\rangle\) and \(|111\rangle\), 
which once more violates any monogamy relation that one can write down, and this ensemble is devoid of any quantum correlations in its element states. 
Note that just like in \textbf{Case I}, all the ensembles are distinguishable by LOCC between all the three parties.

\textbf{Case III.}
The violations in \textbf{Case II} are all obtained for tripartite ensembles. However, this is not a necessary restriction, and ensembles of 
an arbitrary number of parties can be shown to maximally violate the corresponding monogamy, as seen in the following example. 
We present the example by again using a two-element ensemble, where each element is genuinely multiparty entangled \cite{genuine-multi}.  
Let us therefore consider the ensemble \(({\cal E}_{cat})\) containing the GHZ states (also called cat states)
\begin{eqnarray}
 |\psi\rangle^{cat}_{AB_1B_2 \ldots B_N} &=& \frac{1}{\sqrt{2}} (|000 \ldots 0\rangle + |111 \ldots 1\rangle), \nonumber \\
 |\phi\rangle^{cat}_{AB_1B_2 \ldots B_N} &=& \mathbb{I}_A \otimes \sigma^x_{B_1} \otimes \ldots \otimes \sigma^x_{B_N}
 |\psi\rangle^{cat}_{AB_1B_2 \ldots B_N}, \nonumber \\
\end{eqnarray}
where \(\sigma^x\) is the Pauli spin-flip operator. 
Tracing out all the parties except \(A\) and \(B_1\), we obtain the ensemble 
\begin{eqnarray}
 \rho_{\psi^{cat}}^{AB_1} &=& \frac{1}{2}(|00\rangle\langle00| + |11\rangle\langle11|), \nonumber \\
 \rho_{\phi^{cat}}^{AB_1} &=& \frac{1}{2}(|01\rangle\langle01| + |10\rangle\langle10|).
\end{eqnarray}
This ensemble can be exactly distinguished, and so we have 
\begin{equation}
 I_{acc}^{LOCC,A:B_1}({\cal E}_{cat}) = 1.
\end{equation}
However, the ensemble \(({\cal E}_{cat})\) is invariant with respect to a swap between any two of the \(B_i\)s (\(i=1,2,\ldots,N\)), and 
so we have 
\begin{equation}
 I_{acc}^{LOCC,A:B_i}({\cal E}_{cat}) = 1, \quad \forall i = 1,2,\ldots, N.
\end{equation}
However, each of the \(I_{acc}^{LOCC,A:B_i}({\cal E}_{cat})\) can reach a maximum of unity, as \({\cal E}_{cat}\) is a two-element ensemble. 
Therefore, for any monogamy relation to be nontrivial, we must have the sum \(\sum_{i=1}^{N}I_{acc}^{LOCC,A:B_i}\) strictly less than \(N\). 
However, the sum is actually equal to \(N\) for the ensemble \({\cal E}_{cat}\).

\textbf{Case IV.} The ensembles that we have considered until now are all qubit ensembles, and have  two elements in them. Neither of these conditions 
are necessary. For example, the ensemble \({\cal E}_T\), consisting of the three three-qutrit quantum states (shared between \(A\), \(B_1\), and \(B_2\) 
respectively),
%
%
\begin{eqnarray}
&&|000\rangle + |111\rangle, \nonumber \\ 
&&|011\rangle + |122\rangle, \nonumber \\
&&|100\rangle + |200\rangle, 
\end{eqnarray}
also violates any monogamy relation to the maximal extent. 
Also, the ensembles that have been considered above do not form complete bases of the corresponding multiparty Hilbert space. Again, this condition is 
not a necessity, as the ensemble \({\cal E}_P\) formed by the eight three-qubit quantum states \(|ijk\rangle\) (\(i,j,k=0,1\)), form a complete basis, and
also violates the monogamy relation for \(I_{acc}^{LOCC}\). 
It is also not necessary to consider orthogonal ensembles to violate monogamy, as has been done until now. This can be seen by considering 
the ensemble consisting of the states \(|000\rangle\) and \(|\vec{n}\vec{n}\vec{n}\rangle\), where \(|\vec{n}\rangle\) is a qubit 
state slightly different from \(|1\rangle\). 

\textbf{Case V.}
We now consider the ensemble \({\cal E}_{shifts}\) of the following four three-qubit, (globally) orthogonal, quantum states
\begin{equation}
 |01+\rangle, \quad  
 |1+0\rangle, \quad 
 |+01\rangle, \quad 
 |---\rangle, 
\end{equation}
where \(|\pm\rangle = \frac{1}{\sqrt{2}}(|0\rangle \pm |1\rangle)\). 
This set was discovered in Ref. \cite{UPB1}, and it was shown that the set forms an unextendible product basis, 
in the sense that there are no product states in the orthogonal complement of the subspace spanned the elements in \({\cal E}_{shifts}\). It was 
shown there that this ensemble cannot be distinguished by LOCC between the three parties. 
It was thereafter connected to the phenomenon of bound entanglement \cite{bound-ent}. 
We will now show that the ensemble \({\cal E}_{shifts}\) violates monogamy. 
Let the three parties possessing the states in \({\cal E}_{shifts}\) be called \(A\), \(B_1\), and \(B_2\) respectively. 
Leaving out \(B_2\), the ensemble consists of the states
\begin{equation}
|01\rangle, \quad
|1+\rangle, \quad
|+0\rangle, \quad
|--\rangle. 
 \end{equation}
Let us try to find the locally accessible information for this ensemble. Let us begin by noting that the universal upper and lower bounds on locally accessible  information 
imply that 
\begin{equation}
 1-\frac{1}{2}\log_2 \mbox{e} \leq I_{acc}^{LOCC, A:B_1}({\cal E}_{shifts}) \leq 2,
\end{equation}
that is 
\begin{equation}
0.27865 \leq I_{acc}^{LOCC, A:B_1} ({\cal E}_{shifts}) \leq 2.
\end{equation}
Here and hereafter, all numerical values are rounded off to the fifth decimal place.
The upper bound does not help us in violating any monogamy relation. And the lower bound is too weak for our purposes. We now find a much stronger lower bound. 
This is obtained by using the following LOCC measurement strategy, which includes a single bit of 
classical communication from \(A\) to \(B_1\). 
Suppose that \(A\) measures in the basis \(\mathbb{Z} = \{|0\rangle, |1\rangle\}\), and
sends the result to \(B_1\) over a classical channel. If the result is \(|0\rangle\), the observer in possession of \(B_1\) measures in the 
basis \(\mathbb{Z}\), and otherwise he measures in the basis \(\mathbb{X} = \{|+\rangle, |-\rangle\}\).
The mutual information, between the ensemble index and the measurement index, that is obtained by following this LOCC measurement strategy, can be
shown to be 
\begin{eqnarray}
 I_M^{LOCC,A:B_1}({\cal E}_{shifts}) & \geq & \frac{13}{4} - \frac{1}{8}\left[  3 \log_2 3 + 5 \log_2 5 \right] \nonumber \\ 
                                             &  \approx & 1.20443.  
\end{eqnarray}
The bound is much better than the universal lower bound, and indeed will help us to violate monogamy. 
Now
\begin{equation}
 I_{acc}^{LOCC,A:B_1}({\cal E}_{shifts})  \geq  I_M^{LOCC,A:B_1}({\cal E}_{shifts})  \geq 1.20443.  
\end{equation}
Also, the ensemble obtained from \({\cal E}_{shifts}\) by leaving out \(B_1\), is the same as that obtained by leaving out \(B_2\), up to a 
swap operation, and we know that locally accessible information is invariant under the swap operation. Consequently, we have 
\begin{equation}
 I_{acc}^{LOCC,A:B_2}({\cal E}_{shifts})    \geq 1.20443,
\end{equation}
so that
\begin{equation}
 I_{acc}^{LOCC,A:B_1}({\cal E}_{shifts}) +  I_{acc}^{LOCC,A:B_2}({\cal E}_{shifts})    \geq 2.40887. 
\end{equation}
However, since there are four elements in the ensemble \({\cal E}_{shifts}\), we have 
\begin{equation}
 I_{acc}^{LOCC,A:B_1B_2}({\cal E}_{shifts}) \leq 2.
\end{equation}
Consequently, the monogamy relation is violated for the locally indistinguishable ensemble \({\cal E}_{shifts}\) by more than \(20\%\).


In conclusion, we have shown that locally accessible information of multisite quantum ensembles can violate monogamy, even maximally. 
Violation can appear even for locally indistinguishable, but globally orthogonal, multiparty quantum ensembles. This is despite the fact that this 
physically important quantity does satisfy monotonicity under local operations and classical communication.

None of the dual fundamental properties of monogamy and monotonicity are expected to be satisfied by a classical correlation measure of a multiparty system, quantum or 
classical. 
Indeed, just like a single ball can be either green or blue in color, a set \({\cal S}\) of ten (or twenty) balls can be either all green or all blue. 
[Similarly, the spin states of a set of ten spin-1/2 particles can be either all up, in the \(z\)-direction, or all down.] The 
marginals of 
\({\cal S}\) consisting of \emph{any} two balls is again either both green or both blue, irrespective of the number of balls in \({\cal S}\) -- a clear 
violation of monogamy. Also, monotonicity is violated by any classical correlation measure, as can be seen in the following scenario.
Suppose that two white  balls are sent to two cities, so that initially 
there are no correlations in this bipartite system. The receivers of the balls are then  
instructed to color them to a some single color and to choose that color from among green, blue, and red, and to fix the color by phone 
call between the parties. So finally, classical correlation is present, whatever be its value (that depends on the exact measure (and its normalization) used),
in the bipartite system.

While classical correlation measures are not expected to satisfy these dual properties, quantum correlations are. Locally accessible 
information, therefore, contains elements of both the worlds. 

On the application front, the results may have  implications for quantum communication networks, where classical information is transferred by using quantum means. 
Another potential candidate for application is  distributed quantum computing, where efficient transfer of information is vital for a robust and 
competent performance of the system. Yet another possible region of application is in secure information transfer, where quantum 
correlations of quantum ensembles is an important,  though not very well-understood, physical quantity, and the results obtained may provide inputs towards an 
axiomatic formalism of this quantity.

On the fundamental side, we note that the violations are obtained for systems with an arbitrary number of subsystems, and so are valid even for
macroscopic systems. This can have implications for researches in the quantum-to-classical transition: Macroscopic systems, whether classical or quantum, 
have inherent physical quantities that violate monogamy, but  satisfy monotonicity. 
Lastly, the results  may be important to understand the counterintuitive properties obtained in Refs. \cite{nlwe, UPB1, UPB-rest, two-state, morenon}, 
and may indicate that these 
nonintuitive properties are a product of the violation of monogamy of locally accessible information.


\begin{thebibliography}{99}
\bibitem{logbase2} Throughout the paper, we calculate the all the quantities on amounts of
information transfer in bits (binary digits), and this requires the use of logarithm with base 2 in the definition mutual information.
\bibitem{genuine-multi} A pure multiparty quantum state is said to be genuinely multiparty entangled if the state is bipartite entangled across any 
partition of the set of observers into two groups.
\end{thebibliography}
\end{document}